\documentclass[preprintnumbers,prd,twocolumn,nofootinbib,
superscriptaddress,nobibnotes,showpacs]{revtex4}
\usepackage{amsfonts}
\usepackage{mathrsfs}
\usepackage{epsfig}
\usepackage{graphicx}%
\usepackage{dcolumn}
\usepackage{amsmath}

\begin{document}

\title{Unified dark energy and dark matter from a scalar field
  different from quintessence}
\author{Changjun Gao}
\affiliation{The  National Astronomical Observatories, Chinese
Academy of  Sciences, Beijing, 100012, China}
\author{Martin Kunz}
\affiliation{Astronomy Centre, University of Sussex, Brighton
BN1 9QH, United Kingdom}
\affiliation{D\'epartement de Physique Th\'eorique, Universit\'e de
  Gen\`eve, 24 quai Ernest Ansermet, CH-1211 Gen\`eve 4, Switzerland}
\author{Andrew R. Liddle}
\affiliation{Astronomy Centre, University of Sussex, Brighton
BN1 9QH, United Kingdom}
\author{David Parkinson}
\affiliation{Astronomy Centre, University of Sussex, Brighton
BN1 9QH, United Kingdom}
\date{\today}

%%%%%%%%%%%%%%%%%%%%%%%%%%%%%%%%%%%%%%%%%%%%%%%%%%%%%%%%%%%%%%%%%%%%%%%%%%
\begin{abstract}
We explore unification of dark matter and dark energy in a theory
containing a scalar field of non-Lagrangian type, obtained by direct
insertion of a kinetic term into the energy--momentum tensor.  This
scalar is different from quintessence, having an equation of state
between $-1$ and $0$ and a zero sound speed in its rest frame. We
solve the equations of motion for an exponential potential via a
rewriting as an autonomous system, and demonstrate the observational
viability of the scenario, for sufficiently small exponential
potential parameter $\lambda$, by comparison to a compilation of
kinematical cosmological data.
\end{abstract}
%\keywords{}

\maketitle

%%%%%%%%%%%%%%%%%%%%%%%%%%%%%%%%%%%%%%%%%%%%%%%%%%%%%%%%%%%%%%%%%%%%%%%%%%
\section{Introduction}

Dark energy remains a fundamental mystery, both in terms of its
unexpectedly low but non-zero value and because of the apparent
coincidence of its present density being approximately that of other
components. Attempts to address the coincidence problem have been of
two types. One is to invoke the anthropic principle, perhaps at its
most persuasive when coupled with the concept of the string landscape
\cite{landscape}. The second is to permit the dark energy to be a
dynamical entity, and hope to exploit solutions of scaling or tracking
type to remove dependence on initial conditions (for a review see
Ref.~\cite{CST}). It is however fair to say that no compelling
scenario of the second type has been found that is compatible with the
tight present observational constraints on the equation of state
parameter $w$ \cite{Komatsu5yrWMAP}.

In this paper we study the consequences of modeling the dark energy
using a scalar field that is of non-Lagrangian type. The principle
that fundamental physics should derive from a Lagrangian description
is a deep-seated part of modern physics, as powerfully argued for
instance by Durrer and Maartens \cite{ruth:2008}. It is a measure of
the difficulty of the dark energy problem that there have been
several papers that have abandoned this principle, for instance
modeling the dark energy as a phenomenological fluid which exhibits
a particular scaling with the scale factor \cite{cardone:2004} or
Hubble parameter \cite{dvali:2003}, or even allowing a cosmological
constant with an explicit dependence on time \cite{basi:2009}. Our
proposal too is of this general type; we are closer to traditional
quintessence modeling in adopting a scalar field description, but
consider a scalar field that does not emerge from a Lagrangian.

Dropping the Lagrangian assumption is a major step, and in taking such
a step one wishes to be sure that there is significant payback. Our
model offers one such reward --- it permits a unified description of
dark energy and dark matter as due to the single field we consider.
While our proposal is a speculative one, this opportunity is
significant enough to merit study.

Our proposal is not of course the first to seek to unify dark energy
and dark matter into a single material. Discounting those where the
dark energy arises from a constant term in the action, some examples
are as follows. Ref.~\cite{Padmanabhan:2002sh} proposed a tachyon-type
scalar-field Lagrangian, in which the scalar fluid can be broken up
into dark matter and dark energy components. K-essence unification
of dark matter and dark energy has been studied in 
Ref.~\cite{sch:2004}. Staying instead with the canonical Lagrangian,
Ref.~\cite{Arbey:2006it} introduced a complex scalar field with a
mixed potential made of quadratic and exponential terms, which then
mimic dark matter and dark energy respectively. Alternative strands
with similar goals are study of the generalized Chaplygin gas
\cite{BBS} and of barotropic fluid models \cite{baro}.

\section{Modifying the Einstein equations}

The usual Einstein equations are given by
\begin{eqnarray}
G_{\mu\nu}=\kappa^2\left(T_{\mu\nu}+\Lambda g_{\mu\nu}\right)\;,
\end{eqnarray}
where $G_{\mu\nu}$, $T_{\mu\nu}$ and $\Lambda g_{\mu\nu}$ are the
Einstein tensor, energy--momentum tensor and the cosmological constant
term, respectively. $g_{\mu\nu}$ is the metric tensor, $\Lambda$ is
the cosmological constant and $\kappa^2=8\pi G$.  $\Lambda$ is the
simplest version of dark energy, being time independent and
isotropically and homogeneously distributed in space. It suffers from
the coincidence problem, and to address this we wish to allow the
cosmological constant to evolve.

A simple idea, as adopted in quintessence models, is to allow
$\Lambda$ to be a function of some scalar field $\phi$. One cannot
however allow this dependence to be on $\phi$ alone; the Bianchi
identity
\begin{eqnarray}
\nabla^{\mu}G_{\mu\nu}=0\;,
\end{eqnarray}
and the law of energy--momentum conservation,
\begin{eqnarray}
\nabla^{\mu}T_{\mu\nu}=0\;,
\end{eqnarray}
would force
\begin{eqnarray}
\nabla^{\mu}\left[\Lambda\left(\phi\right) g_{\mu\nu}\right]=0\;,
\end{eqnarray}
requiring $\Lambda(\phi)$ to still be constant.

To remedy this it is necessary to incorporate a dynamical term,
depending on $\nabla^{\mu}\phi$, into the equations. For quintessence
this is done by including a canonical kinetic term in the Lagrangian;
$\Lambda(\phi)$ then becomes the scalar field potential and the total
dark energy density includes both potential and kinetic terms. Here we
propose the simplest possible alternative, which is the direct
insertion of a kinetic term into the energy--momentum tensor:
\begin{eqnarray}
G_{\mu\nu}=\kappa^2\left[
T_{\mu\nu}+\Lambda\left(\phi\right)g_{\mu\nu}-
\frac{1}{2}\nabla_{\mu}\phi\nabla_{\nu}\phi\right]
\;.
\end{eqnarray}
Now $\Lambda(\phi)$ is not necessarily a constant. The equation of
motion for the scalar field is then given by
\begin{eqnarray}
\label{eq:eom0}
\nabla^2\phi-2\frac{d\Lambda}{d\phi} +
\frac{\nabla^{\mu}\phi\nabla^{\nu}\phi\cdot   
\left(\nabla_{\mu}\nabla_{\nu}\phi\right)}{\nabla_{\alpha}
\phi\nabla^{\alpha}\phi}=0
\;,
\end{eqnarray}
This equation follows directly from the Einstein equations (plus the
assumption that for the other components $T_{\mu\nu}$ remains
separately conserved), but this form is more convenient.

At first glance this scalar field looks very much like a quintessence
field.  But in fact it is very different from quintessence, and even
different from K-essence \cite{kessence:2000} where the Lagrangian is
written as a general function of $\phi$ and
$\nabla^{\mu}\phi\nabla_{\mu}\phi$. Indeed, it has no Lagrangian
formulation in the framework of K-essence theory.

We can present the proof as follows. In general, the Lagrangian of
K-essence is given by an arbitrary function
\begin{eqnarray}
\mathscr{L}=\mathscr{L}\left(\phi,X\right)\;,
\end{eqnarray}
where $\phi$ is a scalar field and
\begin{eqnarray}
X=-\frac{1}{2}g^{\mu\nu}\partial_{\mu}\phi\partial_{\nu}\phi\;.
\end{eqnarray}
Since we choose the signature of $(-1,+1,+1,+1)$, we always have
$X\geq 0$. Varying this Lagrangian with respect to the metric we
obtain the energy--momentum tensor in the form
\begin{eqnarray}
T_{\mu\nu}=\mathscr{L}_{,X}\nabla_{\mu}\phi\nabla_{\nu}\phi+
\mathscr{L}g_{\mu\nu}\;,
\end{eqnarray}
where $\mathscr{L}_{,X}$ denotes partial derivative of the Lagrangian
with respect to $X$. By identifying it with the energy--momentum tensor
of the scalar
\begin{eqnarray}
T_{\mu\nu}=-\frac{1}{2}\nabla_{\mu}\phi\nabla_{\nu}\phi+
\Lambda\left(\phi\right)g_{\mu\nu}\;,
\end{eqnarray}
we find the corresponding Lagrangian does not exist. This
demonstration is limited to Lagrangians of \mbox{K-essence} form, but
there is no reason to think that a more general Lagrangian, such as
$\mathscr{L}(X,\phi,R_{\mu\nu}\nabla^{\mu}\phi\nabla^{\nu}\phi,\cdots)$
could lead to our equations while retaining the Einstein form of
gravity.

\section{Dynamical equations}

The equations that govern the evolution of the spatially-flat
Friedmann--Robertson--Walker Universe are
\begin{eqnarray}
\label{eq:main}
 &&3H^2=\kappa^2\left[\frac{1}{2}\dot{\phi}^2+\Lambda\left(\phi\right)+\sum
\rho_i\right],\\&&\nonumber
{2\dot{H}}+3H^2=-\kappa^2\left[-\Lambda\left(\phi\right)+\sum
p_i\right],\\&&
\frac{d\rho_i}{dt}+3H\left(\rho_i+p_i\right)=0,  \label{eq:bgcons} 
\end{eqnarray}
where $H=\dot{a}/a$ is the Hubble parameter, $\rho_{i}$ and $p_i$ are
the energy density and pressure of $i$-th matter component (namely,
relativistic matter, baryonic matter, and so on). $a$ is the scale
factor and dot denotes derivative with respect to the physical time
$t$. We have set $a_0=1$ for the present universe.

From the Einstein equations above we obtain the density and pressure
of our scalar
\begin{eqnarray}
\label{eq:dp}
 &&\rho_{\textrm{sca}}=\frac{1}{2}\dot{\phi}^2+\Lambda\left(\phi\right),
\nonumber\\&&
p_{\textrm{sca}}=-\Lambda\left(\phi\right).
\end{eqnarray}
These can be contrasted with the equivalents for quintessence with the
same potential
\begin{eqnarray}
\label{eq:dpq}
 &&\rho_{\textrm{qui}}=\frac{1}{2}\dot{\phi}^2+\Lambda\left(\phi\right),
\nonumber\\&&
p_{\textrm{qui}}=\frac{1}{2}\dot{\phi}^2-\Lambda\left(\phi\right).
\end{eqnarray}

From the expressions of density and pressure, we know quintessence has
the equation of state $-1\leq w_{\textrm{qui}}\leq 1$ for
$\Lambda\geq0$, while the scalar has $-1\leq w_{\textrm{sca}}\leq
0$. From the conservation equation (\ref{eq:bgcons}) we then know that
the density of quintessence scales in the range $a^{-6}$ to $a^{0}$,
while for the scalar the range is restricted to $a^{-3}$ and
$a^{0}$.\footnote{This is the same range accessible to the simplest
  DBI tachyon model \cite{sen:1998,Padmanabhan:2002sh}, but our
  equation of motion differs from that case.} This property suggests
that the scalar may play the role of both dark matter (scaling
approximately as $a^{-3}$) and dark energy (scaling approximately as
$a^{0}$).

From the expressions for the density and pressure we can further
derive the sound speed in the rest-frame of the scalar fields, and
find that they are different as well:
\begin{eqnarray}
\hat{c}_{\textrm{sca}}^2&=&\frac{\partial p/\partial X}{\partial\rho/\partial
X}=0\;,\nonumber\\
\hat{c}_{\textrm{qui}}^2&=&\frac{\partial p/\partial X}{\partial\rho/\partial
X}=1\;,
\end{eqnarray}
where $X=\dot{\phi}^2/2$. This is the case for any potential
$\Lambda(\phi)$. The vanishing of the sound speed allows our scalar
field to cluster gravitationally more easily than quintessence. Since
this is crucial in order to match the cosmic microwave background
(CMB) data, we derive the behavior of the perturbations explicitly
later on, confirming this result.

The equation of motion for $\phi$ can be derived from
Eqs.~(\ref{eq:main}) or Eq.~(\ref{eq:eom0}) as
\begin{eqnarray}
\label{eq:eoms}
\ddot{\phi}+\frac{3}{2}H\dot{\phi}+\frac{d\Lambda}{d\phi}=0\;.
\end{eqnarray}
Compared with the equivalent equation for quintessence, with the same
potential, there is a significant difference: the friction term in the
equation of motion for the scalar is only half that of the
quintessence. So with increasing redshift, the densities of the scalar
field will increase more slowly than quintessence. In fact, if the
potentials are constant or sufficiently flat such that
$d\Lambda/d\phi\simeq0$ , in a kinetic-dominated regime we will have
\begin{eqnarray}
\rho_{\textrm{qui}}\propto a^{-6}\;,
\end{eqnarray}
for quintessence and
\begin{eqnarray}
\rho_{\textrm{sca}}\propto a^{-3}\;,
\end{eqnarray}
for the scalar. The latter is exactly that of cold dark matter.

\section{The evolution of the homogeneous scalar field}

In order to compute the evolution of the scalar field and to check
whether it is compatible with current data sets, we need to
specify a form for the potential $\Lambda(\phi)$. The simplest
choice is just a constant potential,
 $\Lambda=\textrm{const}$ in Eqs.~(\ref{eq:main}).
Then from Eq.~(\ref{eq:eoms}) we have the density
\begin{eqnarray}
\rho_{\textrm{sca}}=\Lambda+\frac{\rho_{{\rm d}0}}{a^{3}}\;.
\end{eqnarray}
where $\rho_{{\rm d}0}$ is a constant which can be interpretted as the
present dark matter density.
This is exactly the $\Lambda \textrm{CDM}$ model, with $\Lambda$
playing the role of dark energy and
$\nabla_{\mu}\phi\nabla_{\nu}\phi/2$ the role of dark matter.

However, taking the potential to be constant is effectively
reintroducing a pure cosmological constant (c.f. the kinetic K-essence
model of Scherrer \cite{sch:2004}), and hence does not represent a
significant step forward in understanding the nature of dark energy,
although our proposal has a novel nature for the dark matter.  We
therefore choose a more general form of the potential,
\begin{equation}
\Lambda(\phi) =V_0 e^{-\kappa\lambda\phi}
\end{equation}
which we will use throughout the remainder of the paper.
Here $V_0$ and $\lambda$ are two
constants. Without loss of generality, we assume $\lambda>0$.
In the limit $\lambda\rightarrow0$ we recover the constant
potential case and therefore the model is continuously connected
with $\Lambda$CDM, at least where the background evolution
is concerned. 

\subsection{Autonomous system of equations}

To study the evolution of the field,
we set up an autonomous system.
The main equations are given by
\begin{eqnarray} \label{eq:main1}
 &&3H^2=\kappa^2\left[\frac{1}{2}\dot{\phi}^2+\Lambda\left(\phi\right)+
\rho_{\rm r}+\rho_{\rm b}\right],\\&&
{2\dot{H}}+3H^2=-\kappa^2\left[-\Lambda\left(\phi\right)+
\frac{1}{3}\rho_{\rm r}\right],\\&&
\ddot{\phi}+\frac{3}{2}H\dot{\phi}+\frac{d\Lambda}{d\phi}=0\;,
\end{eqnarray}
where $\rho_{\rm r}$ and $\rho_{\rm b}$ are the density of radiation and
baryonic matter, respectively.  They have a constant equation of
state equal to $1/3$ and $0$, respectively. Following
Ref.~\cite{cope:1998}, we introduce the following dimensionless
quantities
\begin{eqnarray}
&& x\equiv\frac{\kappa\dot{\phi}}{\sqrt{6}H}\;, \quad
y\equiv\frac{\kappa\sqrt{\Lambda}}{\sqrt{3}H}\;,\nonumber\\&&
\sqrt{\Omega_{\rm b}}\equiv\frac{\kappa\sqrt{\rho_{\rm b}}}{\sqrt{3}H}\;,
\quad \sqrt{\Omega_{\rm r}}\equiv\frac{\kappa\sqrt{\rho_{\rm r}}}{\sqrt{3}H}\;.
\end{eqnarray}
Here $x^2$ and $y^2$ represent the density parameters of the kinetic
and potential terms respectively.
We expect interesting cases to have the scalar field rolling down the
slope of the potential, so since we have assumed $\lambda>0$, we
should have $x>0$. Then the above equations can be written in the
following autonomous form
\begin{eqnarray}
\label{autoquin1} \frac{dx}{d N}&=&
-\frac{3}{2}x+\frac{\sqrt{6}}{2} \lambda y^2\nonumber \\
&& +\frac{3}{2} x
\left[1-y^2+\frac{1}{3}\left(1-x^2-y^2-\Omega_{\rm b}\right)\right], \\
\frac{dy}{d N} &=&
-\frac{\sqrt{6}}{2}\lambda xy \nonumber \\
&& +\frac{3}{2} y
 \left[1-y^2+\frac{1}{3}\left(1-x^2-y^2-\Omega_{\rm b}\right)\right]\;,\\
 \frac{d \Omega_{\rm b}}{d N}&=&
 -{3}\Omega_{\rm b}\left[y^2-\frac{1}{3}\left(1-x^2-y^2-\Omega_{\rm
     b}\right)\right]\,, 
\end{eqnarray}
together with a constraint equation
\begin{equation}
\label{e:28}
x^2+y^2+\Omega_{\rm b}+\Omega_{\rm r}=1\,.
\end{equation}
Here $N\equiv \ln a$. The equation of state $w_\phi$ and the fraction
of the energy density $\Omega_{\phi}$ for the scalar field are
\begin{eqnarray}
w_\phi&\equiv&\frac{p_\phi}{\rho_\phi}=-\frac{y^2}{x^2+y^2}\;,\\
\Omega_{\phi}&\equiv&\frac{\kappa^2\rho_\phi}{3H^2}=x^2+y^2\;.
\end{eqnarray}

\begin{table*}[t]
\begin{center}
\begin{tabular}{|c|c|c|c|c|c|c|c|}
\hline
Name &  $x$ & $y$ & $\sqrt{\Omega_{\rm b}}$ & Existence & Stability &
$\Omega_\phi$  & $w_\phi$ \\
\hline \hline (a) & 0 & 0 & 0 & All $\lambda$& Unstable node
&   0 & -- \\
\hline (b) & $x$ & 0 & $\sqrt{1-x^2}$ &All $\lambda$ & Saddle line
segment for $0\leq x\leq 1$  & $x^2$ & 0  \\
\hline (c) & $\frac{\sqrt{6}}{3}\lambda$ &
$\sqrt{1-\frac{2}{3}\lambda^2}$ & 0& $\lambda^2\leq \frac{3}{2}$ &
Stable node for $\lambda<\frac{\sqrt{6}}{2}$ & 1 &
$-1+\frac{2}{3}\lambda^2$ \\
\hline
\end{tabular}
\end{center}
\caption[crit]{The properties of the critical points for the
  exponential potential given by
  $\Lambda=V_0e^{-\kappa\lambda\phi}$. } \label{crit0}
\end{table*}

\subsection{Observational requirements}

What constraints does a model seeking to explain both dark matter and
dark energy have to satisfy? Normally, observational constraints are
imposed under the assumption of separate dark matter and dark energy
components, but due to the dark degeneracy, first described by Hu and
Eisenstein \cite{hu:1998} and then further explored in
Refs.~\cite{earlydeg1,earlydeg2,mydeg}, gravitational probes alone are
unable to give a unique decomposition and can only impose constraints
on the total dark sector. In Ref.~\cite{martin:2009} we recently
derived the constraints on a combined dark sector fluid from current
kinematical observations in a model-independent way.  These showed
that the total dark sector equation of state must start at or near the
cold dark matter value $w=0$, and then evolve to become negative by
the present following a particular profile. The standard cosmological
model, e.g.\ as in Komatsu et al.\ \cite{Komatsu5yrWMAP}, predicts a
present total dark sector equation of state of about $-0.78$ (the
weighted mean of the dark matter and cosmological constant
contributions), but in fact this value is only weakly constrained
\cite{mort:2009, martin:2009}. The behavior is more tightly
constrained at higher redshifts, where the actual observational data
lies. In addition, a successful unified dark sector model must
reproduce the present dark sector density
$\Omega_{\textrm{dark}}=0.96$.  Our aim will be to test whether our
model can achieve this.

\subsection{Fixed points and phase portraits}

In Table I, we present the properties of the three fixed points for
the exponential potential. The point (a) corresponds to the
radiation-dominated epoch and this point is unstable. The line segment
(b) corresponds to a scalar plus baryon-dominated epoch and it is a
saddle line segment. In this epoch, the scalar field behaves as dust
matter which has the equation of state $w=0$. The point (c)
corresponds to a scalar-dominated epoch. Point (c) is stable and thus
an attractor. In this epoch, the scalar has an equation of state
$w<-1/3$ if $\lambda<1$, and so the Universe accelerates in this
epoch.

\begin{figure}
\includegraphics[width=8cm]{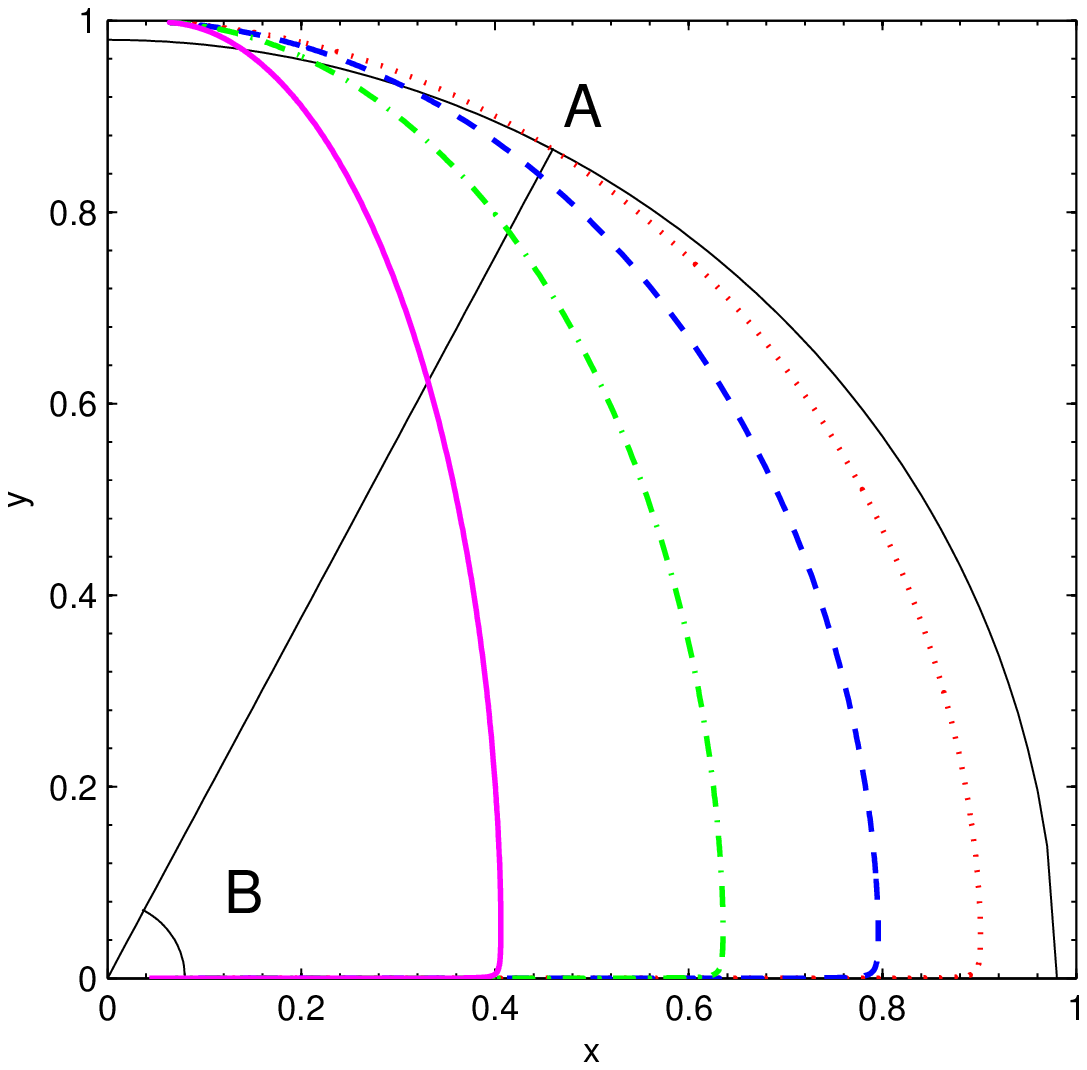}
\caption{The phase plane for the best-fit $\lambda$ found below,
  approximately $0.1$. The point (0,\ 0) corresponds to the
  radiation-dominated epoch. The point (0,\ 0) is unstable and the
  point (0.0082,\ 0.9999) is stable and thus an attractor. The line
  segment($x$,\ 0) is a saddle line. The initial conditions best
  matching observations lead to the red/dotted trajectory, while the
  other trajectories have different initial conditions. The outer thin
  solid line corresponds to $x^2+y^2=1-\Omega_{{\rm b}0}-\Omega_{{\rm
      r}0}$, giving the corrrect present dark sector energy density,
  while the angle B gives the required equation of state $w_\phi =
  -\sin^2 {\rm B} \simeq -0.78$.  So the present-day Universe must lie
  in the vicinity of the point $A$ which is the intersection between
  the two thin solid lines, which the dotted line indeed passes
  through.}\label{fig:expon}
\end{figure}

Viable scenarios start at high redshift near the unstable radiation
fixed point (a). This is necessary since the distance from the origin
corresponds to the relative energy density in the scalar field. As
that energy density, like the one in matter, decreases slower than the
radiation energy density, we need to start close to $x=y=0$, analogous
to the `thawing' regime of quintessence.  In order to follow the usual
evolution of the Universe, the field should then move to the effective
matter-dominated saddle line.  This happens if we start with $y\ll 1$
as $dy/dN\propto y \approx 0$ in this limit. The trajectories then
move across, staying close to the $y=0$ line, before turning
up. Figure~\ref{fig:expon} shows phase portraits for $\lambda=0.01$
with various initial conditions.  The trajectories are all confined
inside the circle given by $x^2+y^2=1$ due to the constraint equation
Eq.~(\ref{e:28}).

The present epoch can be identified through the requirement that
$\Omega_{\phi 0} = x^2 + y^2 \simeq 0.96$, which corresponds to a
circle just inside the limiting circle $x^2+y^2=1$. Since all
trajectories evolve towards the single attractor (c) which lies on
$x^2+y^2=1$, all viable trajectories will cross that line
eventually. In addition, the total equation of state parameter of the
scalar should be of the order of $w_0 \simeq -0.78$ today. That
condition can be graphically represented by a straight radial line
with an angle B $\simeq\arcsin(\sqrt{0.78})$ with respect to the
$x$-axis since $w_\phi = -\sin^2(\textrm{B})$. The good models then
need to cross the circle of today's $\Omega_\phi$ at the intersection
with this line.  We will examine the observational constraints in more
detail in the following subsection.

\begin{figure}
\includegraphics[width=7.5cm]{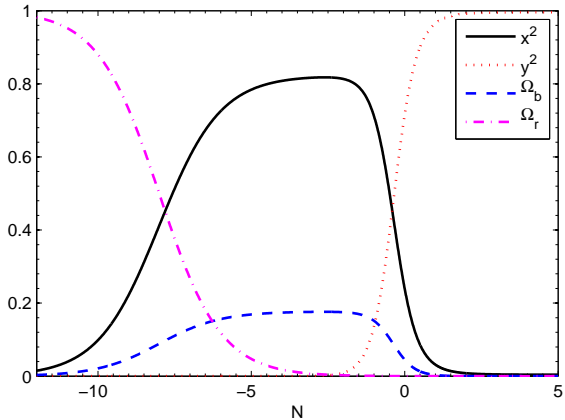}
\caption{The evolution of density fractions for radiation
  (magenta/dot-dashed), baryons (blue/dashed), kinetic term $x^2$
  (black/solid), and potential term $y^2$ (red/dotted), for the
  best-fit model found in subsection~\ref{ss:cons}.}
\label{fig:dens}
\end{figure}

In Fig.~\ref{fig:dens}, we plot the evolution of density fractions for
radiation, baryon matter, kinetic term and potential term, for
best-fit model parameters we determine below.  This shows that the
scalar field can mimic the cold dark matter and dark energy very well.

\subsection{Constraints from current data}

\label{ss:cons}

We now impose detailed observational constraints on our model to
establish its viability.  We follow a method very similar to that
outlined in Ref.~\cite{martin:2009} to constrain the model, applying a
Markov Chain Monte Carlo (MCMC) approach to compute the parameter
posterior probabilities. We assume a flat universe, and fix the
radiation density today from the CMB temperature.  The evolution of
the scalar field is defined in terms of the potential parameter
$\lambda$ and the value of the equation of state today $w_0$, and then
integrated backwards to finds its evolution at earlier time. We assume
a uniform prior on the equation of state parameter of $-1 < w_0 < 0$
and a log prior on $\lambda$ such that $-4 < \log_{10}(\lambda) <
1$. We also include the baryon density $\Omega_{\rm b}h^2$ and the
Hubble parameter today $H_0$ as free parameters, and marginalize over
them.

We use a fairly typical compilation of kinematical data.  Standard
candle data comes from supernova type Ia luminosity distances, for
which we use the cut Union supernova sample \cite{Kowalski:2008ez}
(with systematic errors included), and standard ruler data comes from
the angular positions of the CMB \cite{Wang:2007mza} and Baryon
Acoustic Oscillation peaks \cite{Percival:2009xn}.  Note that
Ref. \cite{Wang:2007mza} gives constraints on the scaled distance to
recombination $R$ and the angular scale of the sound horizon $l_{\rm
  a}$. These are defined to be
\begin{equation}
R \equiv \sqrt{\Omega_{\rm m} H_0^2}\, r(z_{\rm CMB}) \,, \quad l_{\rm
  a} \equiv
\frac{\pi r(z_{\rm CMB})}{r_{\rm s}(z_{\rm CMB})} \,. 
\end{equation}
Since $R$ is scaled by the physical matter density, and so makes
assumptions about the separability of the dark matter and dark energy,
we ignore it in this work. We use only the constraints on $l_{\rm a}$, as
well as those on $\Omega_{\rm b} h^2$ and the correlations between the two.
We also include the SHOES \cite{Riess:2009pu} measurement 
of the Hubble parameter today, 
$H_0=74.2 \pm 3.6 {\rm ~kms}^{-1} {\rm Mpc}^{-1}$. 

\begin{figure}
\includegraphics[width=8.5cm]{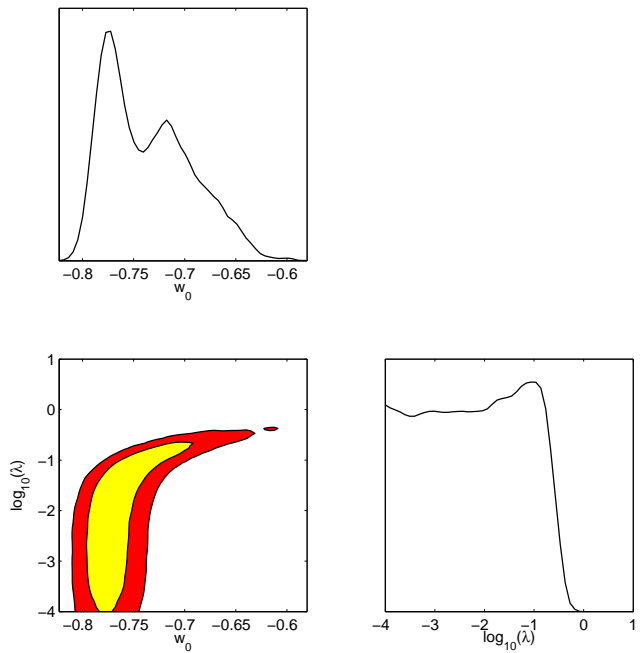}
\caption{The 1-d and 2-d probability distribution for the equation of
  state today, $w_0$, and the potential parameter,
  $\log_{10}(\lambda)$, } \label{fig:2dprobs}
\end{figure}

\begin{figure}
\includegraphics[width=7.5cm]{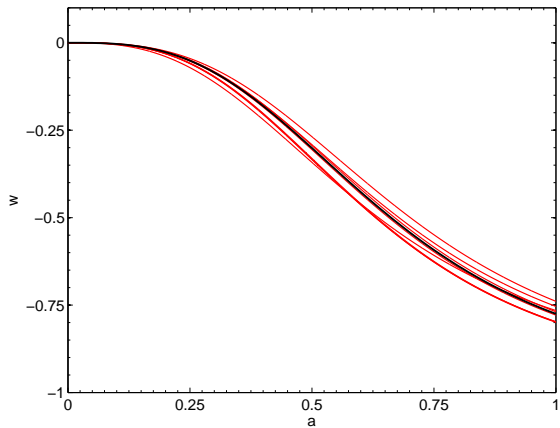}
\caption{The evolution of the equation of state for the scalar, shown
  for 8 models drawn from the Markov chain. It behaves as the cold
  dark matter at higher redshifts and dark energy for the lower
  redshifts. For comparison the evolution of the total $w$ in the
  $\Lambda$CDM model is shown as the thicker black line, which runs
  more or less centrally through the set.} \label{fig:eos}
\end{figure}

We find that the model is a good fit to the data. The best-fit
parameters have a $\chi^2 =312.1$, which is almost equivalent to the
best fit of the LCDM model, $\chi^2=311.9$ (though the scalar field
model has one extra parameter). The equation of state today lies in
the range $-0.82 < w_0 < -0.57$ at 95\% confidence.  The 95\% upper
limit on the potential parameter is $\lambda < 0.20$. The probability
distributions for these two parameters are shown in
Fig.~\ref{fig:2dprobs} and some sample $w(a)$ curves in
Fig.~\ref{fig:eos}. 

\section{Structure formation}

From the analysis in Ref.~\cite{mydeg} we know that the model will fit the
CMB data if the rest-frame sound speed is indeed zero. Because this is
such an important condition on the model, we 
here derive the sound speed directly from the perturbation
equations.  We will discuss the behavior of the perturbations further
in the Appendix.  For this purpose, we work in Newtonian gauge. In the
absence of anisotropic stress, and for scalar perturbations, the
perturbed Friedmann--Robertson--Walker metric can be written in the
form
\begin{eqnarray}
ds^2=-\left(1+2\Phi\right)dt^2+a\left(t\right)^2
\left(1-2\Phi\right)dx_idx^i\;,
\end{eqnarray}
where $\Phi$ is the gauge-invariant Newtonian potential. The potential
characterizes the metric perturbations.

For the dark matter and dark energy dominated Universe, we can safely
neglect the effect of radiation. We assume baryonic matter as a perfect
fluid which has the energy--momentum tensor 
\begin{eqnarray}
T_{\mu\nu}=\left(\rho_{\rm b}+p_{\rm b}\right)u_{\mu}u_{\nu}+p_{\rm b}g_{\mu\nu}\;,
\end{eqnarray}
where $u_{\mu}$ is the four-velocity of the fluid. Perturbations in
the energy density $\rho_{\rm b}$, pressure $p_{\rm b}$ and four-velocity
$u_{\mu}$ can be written as
\begin{eqnarray}
\rho_{\rm b}\left(t,
\vec{x}\right)&=&\rho_{0\rm b}+\delta\rho_{\rm
  b}\left(t,\vec{x}\right)\;,\nonumber\\ 
p_{\rm b}\left(t,
\vec{x}\right)&=&p_{0\rm b}+\delta p_{\rm
  b}\left(t,\vec{x}\right)\;,\nonumber\\ 
u_{\mu}\left(t, \vec{x}\right)&=&{}^{(0)}u_{\mu}+\delta
u_{\mu}\left(t,\vec{x}\right)\;,
\end{eqnarray}
where ${}^{(0)}u_{\mu}=(-1,\ 0,\ 0,\ 0)$ and $\rho_{0\rm
  b}(t),\ p_{0\rm b}(t)$
are the homogeneous and isotropic energy density and pressure. So we
obtain
\begin{eqnarray}
\delta T_0^0&=&\delta\rho_{\rm b}\left(t,\vec{x}\right)\;,\nonumber\\
\delta T_0^i&=&\left(\rho_{0\rm b}+p_{0\rm b}\right)\delta
u^i\left(t,\vec{x}\right)\;,\nonumber\\
\delta T_i^j&=&-\delta p_{\rm b}\left(t,\vec{x}\right)\delta_{i}^{j}\;.
\end{eqnarray}
For the scalar field, we define the perturbation as
\begin{eqnarray}
\phi\left(t,
\vec{x}\right)&=&\phi_0\left(t\right)+\delta\phi\left(t,\vec{x}\right)\;.
\end{eqnarray}
From the energy--momentum tensor
\begin{eqnarray}
T_{\mu\nu}&=&-\frac{1}{2}\nabla_{\mu}\phi\nabla_{\nu}\phi+
\Lambda\left(\phi\right) g_{\mu\nu}\;,
\end{eqnarray}
we get the perturbed counterpart
\begin{eqnarray}
\delta T_0^0&=&\delta \rho_{\phi}=\dot{\phi_0}\dot{\delta
  \phi}-\Phi\dot{\phi_0}^2+\frac{d\Lambda}{d\phi}\delta\phi\;,\nonumber\\
ik\delta T_0^i&=&ik\left(\rho_{\phi0}+p_{\phi0}\right)\delta
u^i_{\phi}\left(t,\vec{x}\right)=\frac{k^2}{2a}\dot{\phi_0}{\delta
  \phi}\equiv{\rho_{\phi0}}V\;,\nonumber\\
\delta T_i^j&=&-\delta
p_{\phi}\left(t,\vec{x}\right)\delta_{i}^{j}=\frac{d\Lambda}{d\phi}\delta\phi
\delta_{i}^{j}\;.
\end{eqnarray}
Since we are working in linear perturbation theory, it is
convenient to transform the equations from real space to Fourier space
since each Fourier mode evolves independently.  We
will also suppress the $0$-subscripts for the homogeneous and
isotropic quantities from now on.

It is well known that both the adiabatic sound speed and the rest
frame sound speed (the sound speed for the fluid in its rest frame)
play a very important role in the discussion of structure formation
theory. Here we work out the two quantities explicitly. The
adiabatic sound speed squared is defined through \cite{martin:2006}
\begin{eqnarray}
c_{\rm a}^2\equiv\frac{\dot{p_{\phi}}}{\dot{\rho_{\phi}}}=
\frac{2}{3H\dot{\phi}}\frac{d\Lambda}{d\phi}\;.
\end{eqnarray}
The rest frame sound speed squared $\hat{c}_{\rm s}^2$ of the scalar is
related to the pressure perturbation in the Newtonian gauge through
\begin{eqnarray}
\label{e:dp} \delta
p_{\phi}=\hat{c}_{\rm s}^2\delta\rho_{\phi}+\frac{3aH}{k^2}
\left(\hat{c}_{\rm s}^2-c_{\rm a}^2\right){\rho_{\phi}}V\;. \label{eq:dpcs}
\end{eqnarray}
Expressing this equation with perturbation quantities, following the
procedure in Ref.~\cite{martin:2006}, we find
\begin{equation}
-\frac{d\Lambda}{d\phi}\delta\phi=
\hat{c}_{\rm s}^2\left(\dot{\phi}\dot{\delta\phi}-\dot{\phi}^2\Phi
+\frac{d\Lambda}{d\phi}\delta\phi+{3H}{\dot{\phi}}\delta\phi\right)
-\frac{d\Lambda}{d\phi}\delta\phi\;.
\end{equation}
Therefore, we can conclude immediately that
\begin{eqnarray}
\hat{c}_{\rm s}^2=0\;.
\end{eqnarray}
This is consistent with the discussion in Eq.~(15). The vanishing
sound speed in the rest frame of the scalar allows it to play the role
of cold dark matter. 

\begin{figure}
\includegraphics[width=7.5cm]{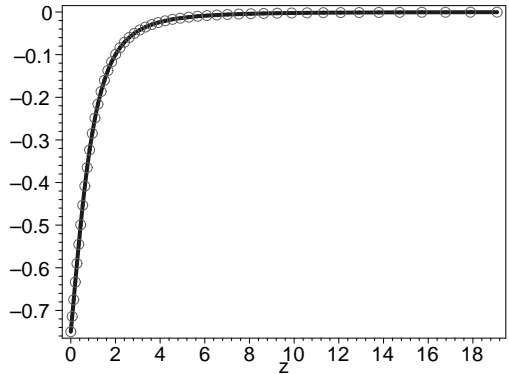}
\caption{The adiabatic sound speed squared $c_{\rm a}^2$ (solid line) and
  the equation of state (circled line) for the scalar, showing
  $c_{\rm a}^2\simeq w_{\phi}$. At redshifts greater than $6$, they are both
  vanishing.}
\label{fig:ssa}
\end{figure}

We have the perturbation for the pressure as follows
\begin{eqnarray}
\delta p_{\phi}=-c_{\rm a}^2\frac{3aH}{k^2}\bar{\rho}V\;.
\label{eq:dpcs0}
\end{eqnarray}
Both CDM and a cosmological constant have $\delta p=0$.
How large is the contribution to $\delta p$ which arises from the
gauge transformation to Newtonian gauge due to $c_{\rm a}^2\neq 0$?
The adiabatic sound speed squared can be written as
\begin{eqnarray}
c_{\rm a}^2=w_{\phi}-\frac{\dot{w_\phi}}{3H\left(1+w_{\phi}\right)}\;.
\label{eq:ca2}
\end{eqnarray}
We note that the adiabatic sound speed is determined by the
homogeneous quantities. To have a picture of the adiabatic sound
speed, we should resort to the background equations, given by 
\begin{eqnarray}
&&\dot{H}=-\frac{\kappa^2}{2}\left(\rho_{\rm
    b}+\frac{1}{2}\dot{\phi}^2\right)\;,\\&& 
\ddot{\phi}+\frac{3}{2}H\dot{\phi}+\frac{d\Lambda}{d\phi}=0\;.
\end{eqnarray}
Using our best-fit parameters, in Fig.~\ref{fig:ssa}, we plot the
equation of state $w_{\phi}$ and the adiabatic sound speed squared
$c_{\rm a}^2$ for the scalar.  It shows that $c_{\rm a}^2\simeq
w_{\phi}$. At the redshifts greater than $6$, both the equation of
state $w_{\phi}$ and the adiabatic sound speed $c_{\rm a}$ are nearly
zero and unimportant for structure formation. This point is essential
for the scalar to play the role of cold dark matter.  In conclusion,
we have $c_{\rm a}\simeq0,\ \ w_{\phi}\simeq0$ at redshifts greater
than about $6$.  Thus we also have $\delta p \simeq 0$ at redshifts
greater than 6, from Eqs.~(\ref{eq:dpcs0}) and (\ref{eq:ca2}).

The Appendix further explores the properties of the structure
formation equations. 

%\begin{eqnarray}
%\frac{\delta
%p_{\phi}}{\delta\rho_{\phi}}=\frac{-\frac{3}{2}H\dot{\phi}\delta\phi
%\left[w_{\phi}-\frac{\dot{w_\phi}}{3H\left(1+w_{\phi}\right)}\right]}
%{\dot{\phi}\dot{\delta\phi}-\dot{\phi}^2\Phi+\frac{d\Lambda}{d\phi}
%\delta\phi}\simeq 0\;,
%|\delta p| = \left | -\frac{d\Lambda}{d\phi} \delta\phi  \right|
%\ll | \delta \rho |  = \left|
%\dot{\phi}\dot{\delta\phi}-\dot{\phi}^2\Phi+\frac{d\Lambda}{d\phi}
%\delta\phi \right| 
%\end{eqnarray}

\section{conclusion}

We have investigated in a cosmological context the behavior of a
scalar field with non-standard kinetic term in the Einstein
equations. So far we lack a Langrangian description for this scalar,
and we have shown that it is not possible to build one in the
framework of K-essence fields. We have not been able to exclude that a
more general Lagrangian, such as
$\mathscr{L}(X,\ \phi,\ R_{\mu\nu}\nabla^{\mu}\phi\nabla^{\nu}\phi,\ \cdot
\cdot\cdot )$ could mimic our equations while retaining the Einstein
form of gravity, or that the Lagrangian looked for could exist in the
framework of Kaluza--Klein theories, but equally we have no reason to
think it will.

On the plus side, we find that this scalar has some interesting
properties. In the first place, it has an equation of state between
$w=-1$ and $w=0$.  This is different from the quintessence field which
has the equation of state between $w=-1$ and \mbox{$w=+1$}. Hence the
scalar field can behave as pressureless matter in the matter- or
radiation-dominated epochs, later evolving to take on dark energy
properties as well. A degree of fine-tuning is needed in the initial
conditions in order to ensure that the scalar field only dominates in
the latter stages of the evolution, which is of the same form as that
invoked in thawing quintessence models.

Secondly, the rest frame sound speed of the scalar is zero. This is
different from quintessence for which the sound speed is equal to the
speed of light.  Although a DBI scalar field has the same range of
equation of state as our scalar, its rest frame sound speed is
nonvanishing. As is known, a vanishing sound speed is sufficient for a
scalar field to play the role of cold dark matter in the process of
structure formation, and the sound speed should not be too big if
gravitational collapse is to match current observations
\cite{mydeg}. Thanks to the vanishing sound speed of our scalar, we
find that it behaves exactly as cold dark matter in the process of
structure formation.

To conclude, we have invoked a scenario in which a non-Lagrangian
scalar field is able to play the roles of both dark matter and dark
energy. With sufficient tuning of initial conditions, we have shown
that a satisfactory evolution can be arranged, with present data
constraining an exponential potential to have an exponent of 0.2 or
less, though the model does not significantly improve on the fit of
the $\Lambda$CDM model. Our model would be falsified in the event of
direct detection of conventional dark matter particles. Whether it can
be motivated in terms of fundamental theories of physics, perhaps in
an effective theory formulation, remains to be seen.

\acknowledgments

The work of C.G.\ is supported by the National Science Foundation of China
under the Distinguished Young Scholar Grant 10525314, the Key
Project Grant 10533010, Grant 10575004, Grant 10973014, the 973
Project (No. 2010CB833004) and the Young Researcher Grant of
National Astronomical Observatories, Chinese Academy of Sciences.
M.K., A.R.L.\ and D.P.\ are supported by STFC (UK). We thank Timothy
Clemson for discussions relating to this paper. C.G.\ acknowledges the
hospitality of the Sussex Astronomy Centre while part of this work was
carried out.

\appendix

\section{Evolution of the perturbations}

In order to make numerical calculations, we should rewrite the
equations in the dimensionless form. To this end, we use the variable
$N$
\begin{eqnarray}
\frac{\partial}{\partial t}=H\frac{\partial}{\partial N}\;,\ \ \
\frac{\partial^2}{\partial t^2}=H^2\frac{\partial^2}{\partial
N^2}+HH'\frac{\partial}{\partial N}\;.
\end{eqnarray}
Here prime denotes the derivative with respect to $N$. So the
background equations become
\begin{eqnarray}
&&HH'=-\frac{\kappa^2}{2}\left(\rho_{\rm
    b}+\frac{1}{2}H^2{\phi'}^2\right)\;,\\&& 
H^2{\phi^{''}}+\frac{1}{2}\left(3H^2+2HH'\right){\phi'}+
\frac{d\Lambda}{d\phi}=0\;.
\end{eqnarray}

Define
\begin{eqnarray}
&&h=\frac{H}{H_0}\;,\ \ \ \Omega_{\lambda}=\frac{\kappa^2
V_0}{3H_0^2}\;,\ \ \ \
\Omega_{{\rm b}0}=\frac{\kappa^2\rho_{{\rm b}0}}{3H_0^2}\;,
\end{eqnarray}
where $\rho_{{\rm b}0},\ \ H_0$ are the energy density of baryon matter and
the Hubble constant in the present-day Universe. Using these new
variables, we can rewrite the main equations in the dimensionless form
\begin{eqnarray}
&&hh'=-\frac{3}{2}\Omega_{{\rm b}0} e^{-3N}-\frac{\kappa^2}{4}
  h^2{\phi'}^2\;,\\&& 
h^2{\phi^{''}}+\frac{h}{2}\left(3h+2h'\right){\phi'}-
3\frac{\lambda}{\kappa}\Omega_{\lambda}e^{-\kappa\lambda\phi}=0\;.
\end{eqnarray}

We can absorb $\kappa$
into $\phi$. Then the main equations are simplified to be
\begin{eqnarray}
&&hh'=-\frac{3}{2}\Omega_{{\rm b}0} e^{-3N}-\frac{1}{4}
h^2{\phi'}^2\;,\\&&
h^2{\phi^{''}}+\frac{h}{2}\left(3h+2h'\right){\phi'}-
3\lambda\Omega_{\lambda}e^{-\lambda\phi}=0\;.
\end{eqnarray}

The perturbed Einstein equations are \cite{ma:1995}
\begin{eqnarray}
3H^2\Phi+3H\dot{\Phi}+\frac{k^2}{a^2}\Phi=-\frac{\kappa^2}{2}\left(\rho_{\rm
  b}\delta_{\rm b}
+\rho_{\phi}\delta_{\phi}\right),\;\;\;\\
k^2H\Phi+k^2\dot{\Phi} = \frac{\kappa^2}{2}
a\left[\left(\rho_{\rm b}+p_{\rm b}\right)\theta_{\rm
    b}+\left(\rho_{\phi}+p_{\phi}\right) 
\theta_{\phi}\right],\;\;\;\\
\ddot{\Phi}+4H\dot{\Phi}+\left(2\dot{H}+3H^2\right)\Phi
=\frac{\kappa^2}{2}\left(-\frac{d\Lambda}{d\phi}\delta\phi\right),\;\;\;\;
\end{eqnarray}
where $\delta_{{\rm b},\phi}\equiv\delta\rho_{{\rm
      b},\phi}/\rho_{{\rm b},\phi}$ is the density contrast for
baryon matter and scalar, respectively, and  $\theta_{{\rm b},\phi}\equiv
i\vec{k}\cdot\vec{v}_{{\rm b},\phi}$ represents the divergence of
velocity for baryon matter and scalar, respectively.  We note that
$\delta_{\phi}$ is different from the perturbation of the scalar,
$\delta\phi$.

On the other hand, the energy conservation equation (which includes
the continuity and Euler equations) holds for the baryon matter and
the scalar field, respectively. So we obtain \cite{ma:1995}
\begin{eqnarray}
\dot{\delta_{\rm b}}&=&-\frac{\theta_{\rm b}}{a}+3\dot{\Phi}\;,\\
\dot{\theta_{\rm b}}&=&-H\theta_{\rm b}+\frac{k^2\Phi}{a}\;,
\end{eqnarray}
for baryon matter \cite{ma:1995}, and
\begin{eqnarray}
\dot{\delta_{\phi}}&=&-\left(1+w_{\phi}\right)\left(\frac{\theta_{\phi}}{a}-
3\dot{\Phi}\right)
\nonumber\\&&-3H\left(\frac{\delta
p_{\phi}}{\delta{\rho_{\phi}}}-w_{\phi}\right)\delta_{\phi}\;,\\
\dot{\theta_{\phi}}&=&-H\left(1-3w_{\phi}\right)\theta_{\phi}-
\frac{\dot{w_{\phi}}}{1+w_{\phi}}\theta_{\phi}
\nonumber\\&&+\frac{\delta
p_{\phi}/\delta{\rho_{\phi}}}{1+w_{\phi}}\frac{k^2}{a}\delta_{\phi}+
\frac{k^2}{a}\Phi\;,
\end{eqnarray}
for the scalar. Here
\begin{eqnarray}
\frac{\delta
p_{\phi}}{\delta\rho_{\phi}}&=&\frac{-\frac{3}{2}H\dot{\phi}\delta\phi
\left[w_{\phi}-\frac{\dot{w_\phi}}{3H\left(1+w_{\phi}\right)}\right]}
{\dot{\phi}\dot{\delta\phi}-\dot{\phi}^2\Phi+
\frac{d\Lambda}{d\phi}\delta\phi}\;,\\
w_{\phi}&=&\frac{-\Lambda}{\frac{1}{2}\dot{\phi}^2+\Lambda}\;.
\end{eqnarray}

The perturbation equation for the scalar is given by
\begin{equation}
\ddot{\delta\phi}+\frac{3}{2}H\dot{\delta\phi}+\frac{k^2}{2a^2}\delta\phi
+2\Phi
\frac{d\Lambda}{d\phi}-\frac{5}{2}\dot{\Phi}\dot{\phi}+
\frac{d^2\Lambda}{d\phi^2}\delta\phi=0\;.
\end{equation}

We find it is convenient to consider the following equations
\begin{eqnarray}
&&\ddot{\Phi}+4H\dot{\Phi}+\left(2\dot{H}+3H^2\right)\Phi=
\frac{\kappa^2}{2}\left(-\frac{d\Lambda}{d\phi}\delta\phi\right),
\;\;\;\;\;\;\\&& 
\ddot{\delta\phi}+\frac{3}{2}H\dot{\delta\phi}+\frac{k^2}{2a^2}\delta\phi
+2\Phi
\frac{d\Lambda}{d\phi}-\frac{5}{2}\dot{\Phi}\dot{\phi}\nonumber \\ &&
\quad+\frac{d^2\Lambda}{d\phi^2}\delta\phi=0\;,\\&&
\dot{\delta_{\phi}}=-\left(1+w_{\phi}\right)\left(\frac{\theta_{\phi}}{a}
-3\dot{\Phi}\right)
\nonumber\\&&
\quad-3H\left(\frac{\delta
p_{\phi}}{\delta{\rho_{\phi}}}-w_{\phi}\right)\delta_{\phi}\;,\\
&&\dot{\theta_{\phi}}=-H\left(1-3w_{\phi}\right)\theta_{\phi}-
\frac{\dot{w_{\phi}}}{1+w_{\phi}}\theta_{\phi}
\nonumber\\&&
\quad +\frac{\frac{\delta
p_{\phi}}{\delta{\rho_{\phi}}}}{1+w_{\phi}}\frac{k^2}{a}\delta_{\phi}
+\frac{k^2}{a}\Phi\;,
\end{eqnarray}
Using the definition of
\begin{eqnarray}
&&\mathfrak{K}=\frac{k}{H_0}\;, \ \ \
{\Theta}_{\phi}=\frac{\theta_{\phi}}{H_0}\;,
\end{eqnarray}
and the variable $N$, we can rewrite the above equations in the
dimensionless form
\begin{eqnarray}
\label{e:pert1}
&&h^2\Phi^{''}+\left(4h^2+hh'\right)\Phi'+\left(2hh'+3h^2\right)
\Phi\nonumber\\&&\quad =\frac{3}{2}\Omega_{\lambda}\lambda
e^{-\lambda\phi}
\delta\phi\;,\\
&&h^2\delta\phi^{''}+\frac{1}{2}\left(3h^2+2hh'\right)\delta\phi'
+\frac{\mathfrak{K}^2}{2a^2}\delta\phi-\frac{5}{2}h^2\Phi'\phi'
\nonumber\\&& \quad -{6\lambda}\Omega_{\lambda}e^{-\lambda\phi}\Phi
+3\Omega_{\lambda}\lambda^2e^{-\lambda\phi}\delta\phi=0\;,
\\&&{\delta_{\phi}}'=-\left(1+w_{\phi}\right)\left(
\frac{\Theta_{\phi}}{ah}-3{\Phi}'\right)
\nonumber\\&&
\quad -3\left(\frac{\delta
p_{\phi}}{\delta{\rho_{\phi}}}-w_{\phi}\right)\delta_{\phi}\;,\\&&
\label{e:pert4}{\Theta_{\phi}}'=-\left(1-3w_{\phi}\right)\Theta_{\phi}-
\frac{{w_{\phi}}'}{1+w_{\phi}}\Theta_{\phi}
\nonumber\\&&\quad +\frac{\frac{\delta
p_{\phi}}{\delta{\rho_{\phi}}}}{1+w_{\phi}}
\frac{\mathfrak{K}^2}{ah}\delta_{\phi}+\frac{\mathfrak{K}^2}{ah}\Phi\;.
\end{eqnarray}
To be consistent with the discussions of the background equations, we have
rescaled $\phi$ by $\phi/\kappa$ as was done earlier.
Correspondingly, we have here
\begin{eqnarray}
&&\frac{\delta
p_{\phi}}{\delta\rho_{\phi}}=\frac{-\frac{3}{2}h^2{\phi}'\delta\phi
\left[w_{\phi}-\frac{{w_\phi}'}{3\left(1+w_{\phi}\right)}\right]}
{h^2{\phi}'{\delta\phi}'-h^2{\phi'}^2\Phi-{3\lambda}
\Omega_{\lambda}e^{-\lambda\phi}\delta\phi}\;,\\&&
w_{\phi}=\frac{-3\Omega_{\lambda}e^{-\lambda\phi}}{\frac{1}{2}h^2{\phi'}^{2}
+3\Omega_{\lambda}e^{-\lambda\phi}}\;.
\end{eqnarray}
Now we have $\Phi$, $\delta\phi$, $\delta_{\phi}$, and
$\Theta_{\phi}$, totalling four
perturbation variables, and four differential equations, namely
Eqs.~(\ref{e:pert1})-(\ref{e:pert4}). Thus the system of equations is
closed.

At redshifts greater than $6$, we have $w_{\phi}\simeq0$, $c_{\rm
  a}^2\simeq0$, and $\delta p_{\phi}/\delta\rho_{\phi}\simeq 0$. So
the perturbation equations simplify to
\begin{eqnarray}
&&h^2\Phi^{''}+\left(4h^2+hh'\right)\Phi'+\left(2hh'+3h^2\right)
\Phi=0\;, \;\;\;\;\;\\
&&{\delta_{\phi}}'=-\left(\frac{\Theta_{\phi}}{ah}-3{\Phi}'\right)
\;,\\&&
{\Theta_{\phi}}'=-\Theta_{\phi}+\frac{\mathfrak{K}^2}{ah}\Phi\;.
\end{eqnarray}
These are none other than the perturbation equations for cold dark
matter in $\Lambda \textrm{CDM}$ model. Therefore, the scalar really
behaves as cold dark matter in the process of cosmic structure
formation.

\newcommand\AL[3]{~Astron. Lett.{\bf ~#1}, #2~ (#3)}
\newcommand\AP[3]{~Astropart. Phys.{\bf ~#1}, #2~ (#3)}
\newcommand\AJ[3]{~Astron. J.{\bf ~#1}, #2~(#3)}
\newcommand\APJ[3]{~Astrophys. J.{\bf ~#1}, #2~ (#3)}
\newcommand\APJL[3]{~Astrophys. J. Lett. {\bf ~#1}, L#2~(#3)}
\newcommand\APJS[3]{~Astrophys. J. Suppl. Ser.{\bf ~#1}, #2~(#3)}
\newcommand\JCAP[3]{~JCAP. {\bf ~#1}, #2~ (#3)}
\newcommand\LRR[3]{~Living Rev. Relativity. {\bf ~#1}, #2~ (#3)}
\newcommand\MNRAS[3]{~Mon. Not. R. Astron. Soc.{\bf ~#1}, #2~(#3)}
\newcommand\MNRASL[3]{~Mon. Not. R. Astron. Soc.{\bf ~#1}, L#2~(#3)}
\newcommand\NPB[3]{~Nucl. Phys. B{\bf ~#1}, #2~(#3)}
\newcommand\PLB[3]{~Phys. Lett. B{\bf ~#1}, #2~(#3)}
\newcommand\PRL[3]{~Phys. Rev. Lett.{\bf ~#1}, #2~(#3)}
\newcommand\PR[3]{~Phys. Rep.{\bf ~#1}, #2~(#3)}
\newcommand\PRD[3]{~Phys. Rev. D{\bf #1}, #2~(#3)}
\newcommand\RMP[3]{~Rev. Mod. Phys.{\bf ~#1}, #2~(#3)}
\newcommand\SJNP[3]{~Sov. J. Nucl. Phys.{\bf ~#1}, #2~(#3)}
\newcommand\ZPC[3]{~Z. Phys. C{\bf ~#1}, #2~(#3)}
\newcommand\JHEP[3]{~JHEP{\bf ~#1}, #2~(#3)}

\end{document}